\documentclass{aa}
\usepackage{psfig}
\usepackage{astron}

\def\m#1{\multicolumn{1}{r@{\mbox{\,\,}}}{#1}} 

\newcommand{\cosec}{\mbox{$\mathrm{cosec}$}}

\begin{document}

\thesaurus{02.19.2; 04.19.1; 08.16.6; 13.18.5}

\title{Pulsars in the Westerbork Northern Sky Survey} 

\titlerunning{Pulsars in the Westerbork Northern Sky Survey}

\author{M.L.A.\ Kouwenhoven}

\authorrunning{M.L.A.\ Kouwenhoven}

\offprints{M.L.A. Kouwenhoven}

\mail{M.L.A.Kouwenhoven@astro.uu.nl}

\institute{   Astronomical Institute,
              P.O.Box 80000, 3508 TA Utrecht, The Netherlands
                }

\date{Received 2000 March 27; accepted May 18}

\maketitle

\begin{abstract}
I have searched the Westerbork Northern Sky Survey (WENSS) source list
for detections of known radio pulsars.  A source with a flux
density greater than five times the local noise level is found near
the positions of 25 pulsars. The probability that one out of these 25
sources is a chance coincidence is about 10\%. I have looked at the
WENSS maps of the non-detected pulsars. A flux density between
three and five times the local noise level is found near the positions
of 14 of these non-detected pulsars. There is a 50 percent probability
that (at least) one of these marginal detections is just a noise
fluctuation. Fourteen radio pulsars, which according to earlier flux
measurements have flux densities above three times the WENSS
noise level, are not detected.  Of the 39 pulsars detected in the
WENSS 19 are also detected in the NRAO VLA Sky Survey (NVSS). By
combining the WENSS and NVSS flux densities for these 19 pulsars
spectral indices are obtained that differ by up to 50\%~ from the long
term averaged values reported in the literature. This affects the
reliability of pulsar candidates that are selected on the basis of
their WENSS-NVSS spectral index.

\end{abstract}

\section{Introduction}

Pulsars can have a steep radio spectrum at frequencies around 1 GHz
and can be highly polarised (e.g.\ Manchester \& Taylor 1977)
\nocite{mt77}. Synthesis maps can therefore be used to select pulsar
candidates on the basis of their steep spectrum and/or high degree of
polarisation. These sources are later observed with a high time
resolution instrument to search for pulsations. At least two pulsars,
PSR~B1937+21 \cite{bkh+82} and
PSR~J0218+4232 \cite{nbf+95}, have been found in this way, while they
were missed in regular pulsar surveys because their pulses were
smeared out in the detection process due to their small period and
their high dispersion measure.

The pulsar population found in this way, may supplement the presently
known population, since this method has totally different selection
effects. These effects can be investigated by studying the spectral
indices and polarisation degrees of known pulsars in these continuum
observations. The spectral indices are strongly influenced by
scintillation. This may cause the flux density of a pulsar to
vary by more than 100 percent on time scales of minutes (diffractive
scintillation) to days (refractive scintillation).

In this paper I search for detections of pulsars in the
Westerbork Northern Sky Survey (WENSS), a survey performed at 325
MHz. I compare my results with those from Kaplan et al.\
\cite*{kcac98} and Han \& Tian \cite*{ht99}, who did similar analyses
with data from the 1400 MHz NRAO VLA Sky Survey (NVSS).  Sect.\
\ref{sec:wenss} describes the WENSS and Sect.\ \ref{sec:correlation}
describes how the pulsar catalog was correlated with the WENSS source
catalog. In Sect.\ \ref{sec:marginal} the positions of the
non-detected pulsars are searched for flux densities above three
times the local noise level. Sect.\ \ref{sec:nondetections} discusses
the remaining non-correlations. Sect.\ \ref{sec:specind} combines the
WENSS results with the NVSS correlation studies to determine the
spectral indices and compares these with values reported in other
literature. Finally, in Sect.\ \ref{sec:discussion} the role of
scintillation is discussed.

\section{WENSS}
\label{sec:wenss}

The Westerbork Synthesis Radio Telescope (WSRT) is an east-west array,
consisting of fourteen 25 meter dishes \cite{bh74}. A twelve hour
observation is needed to get spatial resolution in all directions. Ten
of the dishes have fixed positions and are spaced 144 meters apart.
The remaining four are moveable, although their mutual distances are
usually kept constant. By observing an object several times with
different distances between the fixed and moveable subarray, different
baselines are obtained and the $uv$-coverage improves, yielding an
improved synthesised antenna pattern.

The Westerbork Northern Sky Survey (WENSS) is a 325 MHz continuum
survey of the sky above declination +30\degr (Rengelink et al.\ 1997,
de Bruyn et al.\ in preparation). This area was surveyed using a
mosaicing technique. Each field was observed 18 times for 20 seconds
spread over 12 hours. Each mosaic was observed on six days with a
different spacing between the fixed and moveable subarray to get a
uniform spatial distribution. These observations were spread over
periods of weeks to years. The resulting flux densities are
averaged over all these observations.

\nocite{rtb+97}

The WENSS beam size is 54\arcsec $\times$ 54\arcsec~$\cosec ~\delta$
(FWHM). The final maps have pixel sizes of 21.09\arcsec. When a pixel
was found with a flux density above five times the local noise
level, a two-dimensional Gaussian was fitted to its surroundings. The
coordinates of the centroid of the fit, the maximum (peak) flux
density and the flux density integrated over the fit were
added to the source list. Extra flags in the catalog mark multiple and
extended sources. For a point source the peak flux density
equals the flux density integrated over the beam. Only the peak
flux densities are used in this analysis, since pulsars are
intrinsically point sources. Bright sources have positional errors of
1.5\arcsec, weaker sources at lower declinations have errors up
to about 10\arcsec. On average the uncertainty is 5\arcsec.

\begin{figure}
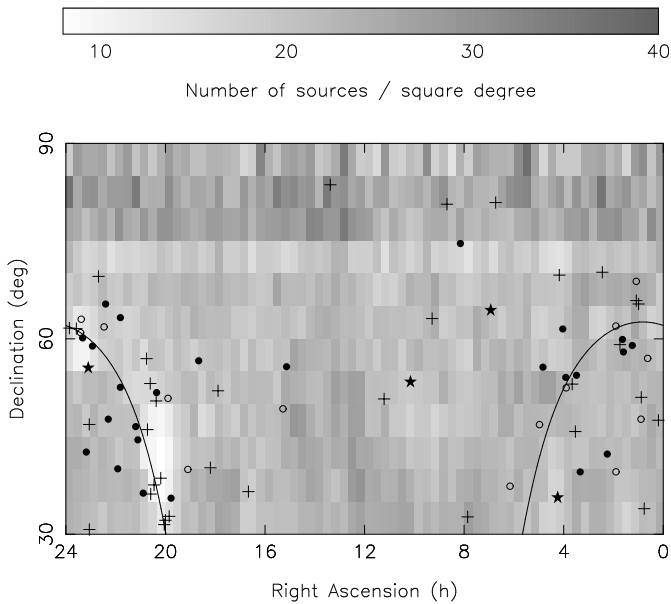

\centerline{\psfig{figure=positions_scale.ps,bbllx=491pt,bblly=33pt,bburx=383pt,bbury=750pt,angle=-90,width=\columnwidth,clip=t}}
\vspace*{4mm}
\centerline{\psfig{figure=positions.ps,angle=-90,width=\columnwidth,clip=t}}
\caption{Pulsar positions and WENSS source densities. Filled circles
mark the positions of the pulsars which are correlated with a WENSS
source. Open circles indicate the pulsars, with a marginal detection
in the WENSS. Confused pulsars are marked with a star. Pulsars with no
WENSS counterpart are indicated with a plus.  The grey-scale indicates
the WENSS source density (white is low density). The polar cap
region has been surveyed with a larger bandwidth and has a higher
source density. A small region near
Cygnus A (RA 20h, Dec 40$^\circ$) is blanked in the WENSS. The solid
line indicates the Galactic plane.}
\label{fig:pos}
\end{figure}

The total bandwidth was 5 MHz.  For most regions of the sky the
detection limit was between 15 and 25 mJy (five times the local noise
level). The polar cap (declination $>$ +75\degr) was surveyed with a
larger bandwidth and the detection limit for this area was about 10 --
12 mJy. The WENSS source catalog contains 229420 sources
\cite{b+98}. A total of 18186 of these are located in the polar
cap area. Fig.\ \ref{fig:pos} shows the density of sources in the
WENSS area.

\section{Detected pulsars}
\label{sec:correlation}

\subsection{Positional correlation}

The radio pulsar catalog (Taylor et al., 1993, Taylor et al., 1995)
contains 84 entries in the part of the sky that was covered by the
WENSS area. The typical positional uncertainty for a pulsar with
a flux density greater than 10 mJy is 0.1\arcsec~or less. In most cases
the uncertainty in the pulsar position is negligible compared to the
positional uncertainty of WENSS sources.

\nocite{tml93}
\nocite{tmlc95}

The pulsar proper motions are neglected, since for each pulsar the
change in position between the epoch of discovery and the epoch of the
WENSS is less than 0.5\arcsec~in right ascension and less than
0.8\arcsec~in declination for all these pulsars. This is much smaller
than the uncertainties in the WENSS positions.

Seven pulsars in the WENSS area have large positional errors of
about 4\arcmin. The probability that a WENSS source is located by
chance within a circle of three times this positional error is more
than 90 percent, if it assumed that the WENSS sources are uniformly
distributed over the sky. Therefore, I have excluded these 7 pulsars
(PSRs J0417+35, B1639+36B, J1758+30, J1900+30, J1931+30, J2002+30 and
J2304+60) from further analysis. PSR~B2000+40 was also excluded,
since it is located in the Cygnus A region, where no WENSS map could
be made.

I have taken the J2000 positions of the remaining 76 pulsars and
compared them with the positions of the sources in the WENSS catalog.
Twenty-five pulsars have a WENSS source located within three times
their combined positional uncertainty $\sigma$, with
\begin{equation}
  \sigma = \sqrt{ \left( \frac{\Delta\alpha}{\sigma_\alpha} \right)^2
             + \left( \frac{\Delta\delta}{\sigma_\delta} \right)^2 },
\end{equation}
where $\Delta\alpha$ and $\Delta\delta$ are the positional differences
in right ascension and declination, respectively, and
\begin{equation}
  \sigma_\alpha = \sqrt{\sigma_{\alpha,{\rm WENSS}}^2 +
    \sigma_{\alpha,{\rm PSR}}^2}
\end{equation}
and similarly for $\sigma_\delta$. Table \ref{tab:pos} lists the J2000
positions of the correlated pulsars and WENSS sources and their
offset, both in arcseconds and in $\sigma$.

In Fig.\ \ref{fig:pos} it can be seen that the pulsars are located in
areas with different values for the WENSS source density. The
probability of a change coincidence can be approximated by a
probability calculation that assumes a uniform distribution of the
WENSS sources. In that case, the probability that an individual
correlation is just by chance is 0.0012.  The binomial probability,
that one out of 76 trials gives a chance correlation is 0.083. The
probability that two correlations occur by chance is 0.004.

\begin{table*}
\caption{J2000 positions of pulsars and correlated WENSS
sources. Offsets are the roots of the sum of the squared differences
in right ascension and declination in \arcsec~(absolute) and in
$\sigma$ (relative).}
\label{tab:pos}
 \begin{tabular}{lc@{ }c@{ }c@{.}c@{ $\pm$ }r@{.}l
 c@{ }c@{ }c@{.}c@{ $\pm$ }r@{.}lc@{ }c@{ }c@{.}c@{ $\pm$ }r@{.}l
 c@{ }c@{ }c@{ $\pm$ }rr@{.}lr@{.}l}
 \hline\hline
 \multicolumn{1}{c}{PSR} & 
 \multicolumn{12}{c}{PSR Position (J2000)} & 
 \multicolumn{10}{c}{WENSS Position (J2000)} &
 \multicolumn{4}{c}{Offset}  \\
 \multicolumn{1}{c}{Name} & 
 \multicolumn{6}{c}{R.A.} & 
 \multicolumn{6}{c}{Dec.} & 
 \multicolumn{6}{c}{R.A.} & 
 \multicolumn{4}{c}{Dec.} & 
 \multicolumn{2}{c}{abs} & 
 \multicolumn{2}{c}{rel}  \\
  & h & m & \multicolumn{2}{c}{s~~~~~~} & 
  \multicolumn{2}{c}{ } & 
  $^\circ$ & $^\prime$ & 
  \multicolumn{2}{c}{$^{\prime\prime}$~~~~~} &
  \multicolumn{2}{c}{ } & 
  h & m & \multicolumn{2}{c}{s~~~~~} & 
  \multicolumn{2}{c}{ } & 
  $^\circ$ & $^\prime$ & 
  \multicolumn{1}{c}{$^{\prime\prime}$~~~~} & 
  \multicolumn{1}{c}{ } & 
  \multicolumn{2}{r}{$^{\prime\prime}$~~~} & 
  \multicolumn{2}{c}{$\sigma$} 
  \\ \hline 
  ~\\
B0114+58   & 01 & 17 & 38 & 702 &   0 & 007 & 59 & 14 & 37 & 85 &    0 & 07 &
01 & 17 & 38 & 3 & 0 & 9 & 59 & 14 & 39 &  8 &     3 & 18 &     0 & 47 
  \\
B0136+57   & 01 & 39 & 19 & 770 &   0 & 003 & 58 & 14 & 31 & 85 &    0 & 03 &
01 & 39 & 19 & 7 & 0 & 3 & 58 & 14 & 32 &  3 &     0 & 81 &     0 & 30 
  \\
B0138+59   & 01 & 41 & 39 & 947 &   0 & 007 & 60 & 09 & 32 & 28 &    0 & 05 &
01 & 41 & 39 & 6 & 0 & 4 & 60 & 09 & 36 &  4 &     4 & 83 &     1 & 38 
  \\
J0218+4232 & 02 & 18 & 06 & 350 &   0 & 010 & 42 & 32 & 17 & 50 &    0 & 10 &
02 & 18 & 06 & 5 & 0 & 2 & 42 & 32 & 16 &  3 &     1 & 87 &     0 & 78 
  \\
B0320+39   & 03 & 23 & 26 & 605 &   0 & 006 & 39 & 44 & 53 & 06 &    0 & 07 &   
03 & 23 & 26 & 4 & 0 & 4 & 39 & 44 & 58 &  8 &     5 & 20 &     0 & 78 
  \\
       ~\\
B0329+54   & 03 & 32 & 59 & 347 &   0 & 010 & 54 & 34 & 43 & 25 &    0 & 10 &
03 & 32 & 59 & 4 & 0 & 2 & 54 & 34 & 45 &  2 &     1 & 38 &     0 & 91 
  \\
B0355+54   & 03 & 58 & 53 & 705 &   0 & 004 & 54 & 13 & 13 & 58 &    0 & 03 &
03 & 58 & 54 & 1 & 0 & 3 & 54 & 13 & 15 &  3 &     4 & 04 &     1 & 73 
  \\
B0402+61   & 04 & 06 & 30 & 052 &   0 & 014 & 61 & 38 & 40 & 76 &    0 & 16 &
04 & 06 & 30 & 8 & 1 & 1 & 61 & 38 & 39 &  9 &     5 & 75 &     0 & 70 
  \\
B0450+55   & 04 & 54 & 07 & 621 &   0 & 003 & 55 & 43 & 41 & 22 &    0 & 10 &
04 & 54 & 08 & 3 & 0 & 3 & 55 & 43 & 44 &  3 &     6 & 44 &     2 & 32 
  \\
B0809+74   & 08 & 14 & 59 & 443 &   0 & 040 & 74 & 29 & 05 & 79 &    0 & 14 &
08 & 14 & 59 & 2 & 0 & 6 & 74 & 29 & 05 &  2 &     1 & 62 &     0 & 72 
  \\
       ~\\ 
B1508+55   & 15 & 09 & 25 & 724 &   0 & 009 & 55 & 31 & 33 & 01 &    0 & 08 &
15 & 09 & 25 & 7 & 0 & 2 & 55 & 31 & 33 &  2 &     0 & 53 &     0 & 28 
  \\
B1839+56   & 18 & 40 & 44 & 594 &   0 & 050 & 56 & 40 & 55 & 58 &    0 & 40 &
18 & 40 & 45 & 5 & 0 & 9 & 56 & 40 & 49 &  8 &     9 & 89 &     1 & 32 
  \\
B1946+35   & 19 & 48 & 25 & 037 &   0 & 002 & 35 & 40 & 11 & 28 &    0 & 02 &
19 & 48 & 25 & 1 & 0 & 2 & 35 & 40 & 09 &  2 &     2 & 35 &     1 & 01 
  \\
B2021+51   & 20 & 22 & 49 & 900 &   0 & 002 & 51 & 54 & 50 & 06 &    0 & 02 &   
20 & 22 & 49 & 4 & 0 & 4 & 51 & 54 & 48 &  5 &     4 & 61 &     1 & 15 
  \\
B2053+36   & 20 & 55 & 31 & 343 &   0 & 040 & 36 & 30 & 21 & 13 &    0 & 50 &   
20 & 55 & 31 & 3 & 0 & 3 & 36 & 30 & 12 &  6 &     9 & 28 &     1 & 46 
  \\
       ~\\
B2106+44   & 21 & 08 & 20 & 478 &   0 & 010 & 44 & 41 & 48 & 79 &    0 & 10 &
21 & 08 & 21 & 0 & 0 & 6 & 44 & 41 & 41 &  8 &     9 & 31 &     1 & 29 
  \\
B2111+46   & 21 & 13 & 24 & 295 &   0 & 014 & 46 & 44 & 08 & 68 &    0 & 11 &
21 & 13 & 24 & 3 & 0 & 2 & 46 & 44 & 08 &  2 &     0 & 55 &     0 & 28 
  \\
B2148+63   & 21 & 49 & 58 & 594 &   0 & 030 & 63 & 29 & 43 & 53 &    0 & 20 &
21 & 49 & 59 & 9 & 0 & 7 & 63 & 29 & 51 &  5 &    11 & 66 &     2 & 31 
  \\
B2148+52   & 21 & 50 & 37 & 742 &   0 & 007 & 52 & 47 & 49 & 67 &    0 & 05 &
21 & 50 & 38 & 2 & 0 & 7 & 52 & 47 & 54 &  8 &     6 & 21 &     0 & 89 
  \\
B2154+40   & 21 & 57 & 01 & 821 &   0 & 013 & 40 & 17 & 45 & 88 &    0 & 14 &
21 & 57 & 01 & 7 & 0 & 2 & 40 & 17 & 46 &  2 &     0 & 95 &     0 & 48 
  \\
       ~\\ 
B2217+47   & 22 & 19 & 48 & 136 &   0 & 004 & 47 & 54 & 53 & 83 &    0 & 04 &
22 & 19 & 48 & 3 & 0 & 2 & 47 & 54 & 54 &  2 &     1 & 25 &     0 & 76 
  \\
B2224+65   & 22 & 25 & 52 & 360 &   0 & 020 & 65 & 35 & 33 & 78 &    0 & 12 &
22 & 25 & 51 & 3 & 1 & 1 & 65 & 35 & 24 &  7 &    12 & 13 &     1 & 68 
  \\
B2255+58   & 22 & 57 & 57 & 711 &   0 & 004 & 59 & 09 & 14 & 95 &    0 & 03 &
22 & 57 & 58 & 0 & 1 & 0 & 59 & 09 & 12 &  9 &     4 & 12 &     0 & 50 
  \\
B2310+42   & 23 & 13 & 08 & 571 &   0 & 006 & 42 & 53 & 12 & 98 &    0 & 03 &
23 & 13 & 08 & 6 & 0 & 2 & 42 & 53 & 13 &  3 &     0 & 43 &     0 & 21 
  \\
B2319+60   & 23 & 21 & 55 & 187 &   0 & 040 & 60 & 24 & 30 & 66 &    0 & 30 &   
23 & 21 & 54 & 4 & 0 & 8 & 60 & 24 & 36 &  7 &     7 & 94 &     1 & 19 
  \\
  ~\\
 \hline\hline
 \end{tabular}

\end{table*}

\begin{figure}
\centerline{\psfig{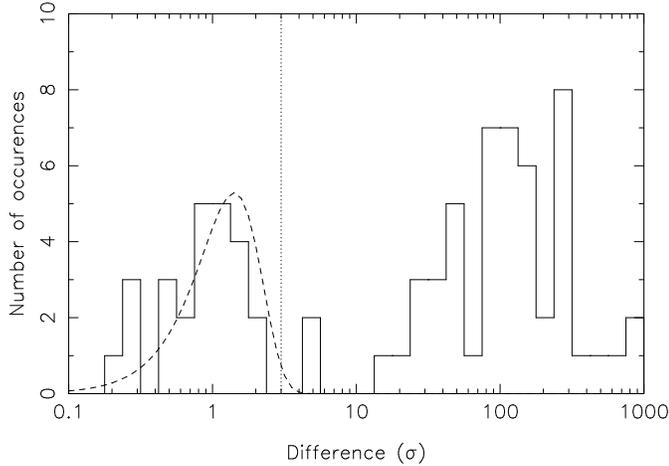}}
\caption{Histogram of differences between the position of a pulsar and
its nearest WENSS source in units of the combined positional
uncertainty $\sigma$. The dotted line markes the 3$\sigma$ limit. The
dashed lines is the expected distribution for 25 correlations,
i.e.\ the number of found pairs with a positional difference below
3$\sigma$.}
\label{fig:posdif}
\end{figure}

Fig.\ \ref{fig:posdif} displays the distribution of positional
differences between a pulsar and its nearest WENSS source in units of
their combined positional uncertainty $\sigma$. There is a clear gap
between the correlated pairs (difference less than 3$\sigma$) and the
non-correlated ones. The distribution of the positional difference
($\Delta$) for the related pairs is
\begin{equation}
  P(\Delta) = \Delta \cdot e^{-\Delta^2/2},
\label{diff_distr}
\end{equation}
which follows after a conversion to polar coordinates. This
distribution is also plotted in Fig.\ \ref{fig:posdif}. A
Kolmogorov-Smirnov test assigns a 15 procent probability that our
sample is drawn from the distribution (\ref{diff_distr}).

There are two objects with positional differences between 3 and 10
$\sigma$. These are in confused regions and will be discussed in
Sect.\ \ref{sec:marginal}.

\subsection{Flux densities}
\label{sec:flux}

\begin{table}
\caption{List of estimated pulsar flux densities (PSR $S_{325}$)
and WENSS source peak flux densities (WENSS $S_{325}$) for the
25 correlated objects.  Navarro et al.\ (1995) observed that the
continuum flux density for PSR~J0218+4232 varies significantly
between 100 and 200 mJy.}
\label{tab:flux}
\begin{tabular}{lr@{.}l@{ $\pm$ }r@{.}lr@{ $\pm$ }rr}
\hline\hline
\multicolumn{1}{c}{PSR} &
\multicolumn{4}{c}{PSR $S_{325}$} &
\multicolumn{2}{c}{$S_{325}$} &
\multicolumn{1}{c}{$S_{\rm noise}$}
\\
\multicolumn{1}{c}{name} &
\multicolumn{4}{c}{estimate} &
\multicolumn{2}{c}{WENSS} 
\\
\multicolumn{1}{c}{ } &
\multicolumn{4}{c}{(mJy)} &
\multicolumn{2}{c}{(mJy)} &
\multicolumn{1}{c}{(mJy)} \\
%
%
\hline
~\\
B0114+58  &  13&8 &   1&6  &   25 &    5   & 3.9 \\
B0136+57  &  36&7 &   2&5  &   70 &    6   & 3.7 \\
B0138+59  & \m{74}&&\m{9}& &   51 &    5   & 3.5 \\
J0218+4232&\m{150}&&\m{50}& &  113 &    7   & 4.4 \\ 
B0320+39  &\m{57}&&\m{12}& &   33 &    5   & 3.9 \\
~\\
B0329+54  &\m{2510}&&\m{380}&&1075&   43   & 5.0 \\
B0355+54  &\m{60}&&\m{10}& &   81 &    5   & 3.4 \\
B0402+61  &  22&5 &   1&5  &   27 &    7   & 5.2 \\
B0450+55  &\m{70}&&\m{11}& &   65 &    5   & 3.5 \\
B0809+74  &\m{122}&&\m{29}&&  213 &   14   & 8.4 \\
~\\
B1508+55  &\m{184}&&\m{11}&&  133 &    7   & 3.0 \\
B1839+56  &\m{31}&& \m{5}& &   23 &    5   & 3.8 \\
B1946+35  &\m{307}&&\m{15}&&  251 &   13   & 6.8 \\
B2021+51  &\m{102}&&\m{22}&&   55 &    7   & 4.8 \\
B2053+36  &  45&5 &   1&7  &   45 &    6   & 4.0 \\
~\\
B2106+44  &  41&5 &   2&6  &   32 &    6   & 4.4 \\
B2111+46  &\m{366}&&\m{21}&&  174 &    9   & 4.0 \\
B2148+63  &  50&6 &   3&2  &  37  &    6   & 4.1 \\
B2148+52  &  19&4 &   1&1  &  26  &    5   & 3.8 \\
B2154+40  &\m{142}&&\m{10}&& 120  &    7   & 3.6 \\
~\\
B2217+47  &\m{210}&&\m{41}&& 216  &   10   & 3.5 \\
B2224+65  &  34&5 &   3&7  &  25  &    5   & 4.0 \\
B2255+58  &  52&7 &   2&8  &  45  &   11   & 8.0 \\
B2310+42  &\m{122}&&\m{16}&& 108  &    6   & 3.6 \\
B2319+60  &\m{71}&& \m{8}& &  81  &   16   & 11.9\\
~\\
\hline\hline
\end{tabular}

\end{table}
\nocite{nbf+95}

Table \ref{tab:flux} lists the observed flux densities of the
correlated WENSS sources and their uncertainties.  The flux
densities can be compared with known pulsar flux densities, but
these are usually measured at frequencies of 400 MHz and higher. By
assuming a power law with constant spectral index, these flux
density data can be extrapolated. I have used flux density
data from Lorimer et al.\ \cite*{lylg95}, hereafter LYLG. These
flux densities are averaged over many observations spread over
years and their uncertainties include variations due to
scintillation. LYLG provide data for 24 of the 25 pulsars in Table
\ref{tab:pos}. Flux densities for PSR~J0218+4232 are taken from
its discovery paper \cite{nbf+95}.

\begin{figure*}
\centerline{\psfig{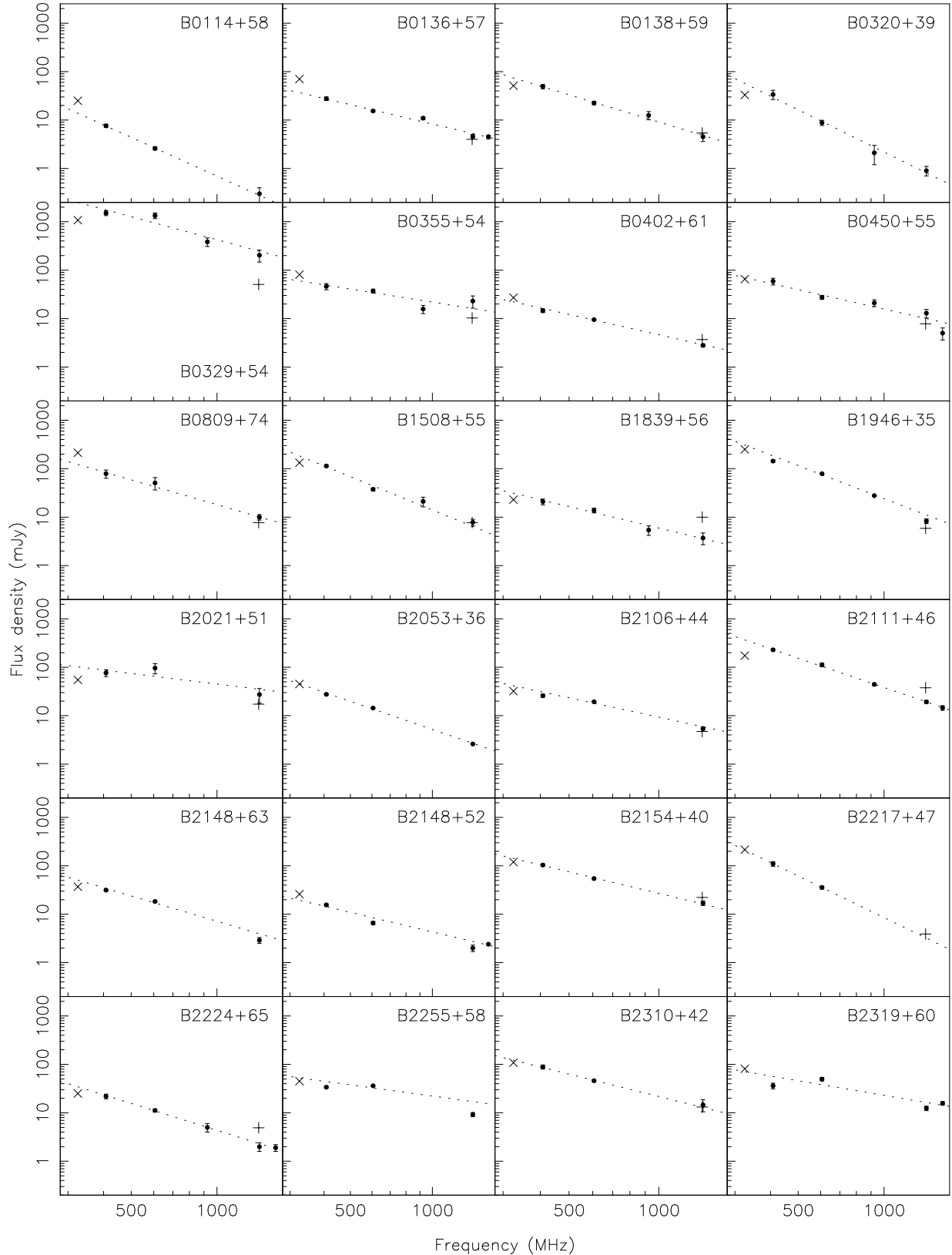}}
\caption{Flux densities against frequency for 24 pulsars with a
WENSS counterpart (PSR~J0218+4232 is excluded, see text).  Filled
circles mark flux densities measured by Lorimer et al.\
(1995). The dotted line is a fitted power law with a constant spectral
index. Flux densities from the corresponding WENSS sources are
indicated by a cross. Flux densities from the corresponding NVSS
sources are marked with a plus. The errors in the WENSS and NVSS
source flux densities are smaller than or of the order of the
size of the symbol and do not include any uncertainty caused by
scintillation.}
\label{fig:si}
\end{figure*}

\begin{figure}
\centerline{\psfig{figure=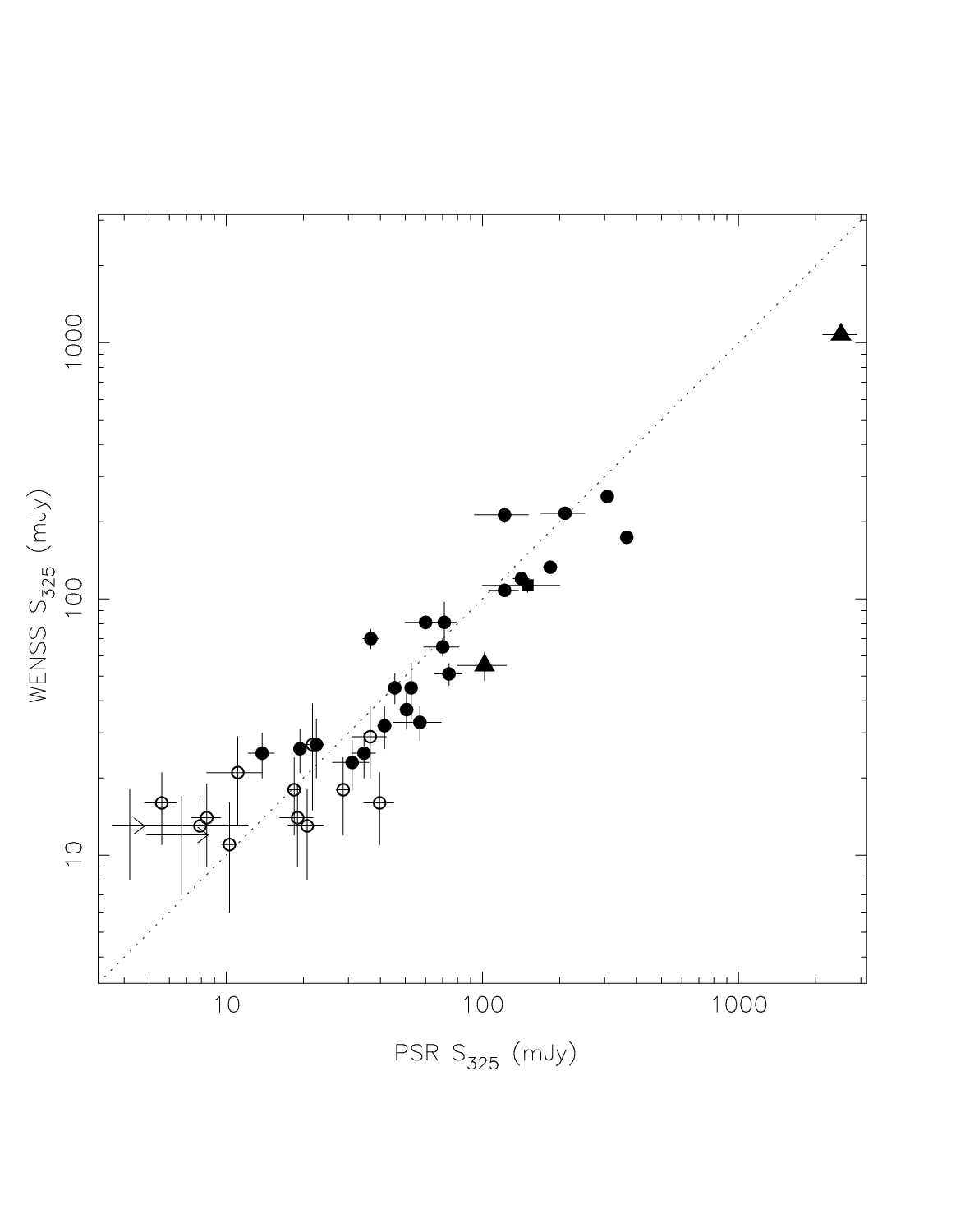,angle=0,width=\columnwidth,bbllx=37pt,bblly=117pt,bburx=538pt,bbury=623pt,clip=t}}
\caption{WENSS flux density as a function of the extrapolated
pulsar flux density (PSR S$_{325}$) for the 25 correlated objects
(filled symbols) and for the marginal objects (clear symbols).
PSR~J0218+4232 is marked with a square. PSR~B0329+54 and
PSR~B2021+51 show indications for a low frequency turnover and are
marked with a triangle.}
\label{fig:flux-WENSS}
\end{figure}

Pulsar flux densities usually obey a power law with a negative
exponent ($S \propto \nu^\alpha$ with $\alpha < 0$) in the
frequency range from 325 to 1400 MHz. I have fitted the logarithms of the
flux densities with a straight line. These lines are plotted in Fig.\
\ref{fig:si}.  From this fit a flux density at 325 MHz is estimated. This
estimate is plotted against the flux density of the WENSS counterpart in Fig.\
\ref{fig:flux-WENSS}. It is known that some pulsars have a low
frequency turnover, usually located around 100 MHz
\cite{mgj+94}. However, some pulsars exhibit a turnover at a higher
frequency: PSR~B0329+54 around 300 MHz \cite{lr68} and PSR~B2021+51
around 400 MHz. The spectrum of PSR B2319+60 is flat between 200 and
600 MHz \cite{mgj+94}. PSRs B1946+35 and B2154+40 also have a
turnover, but it is not clear whether this located between 325 and 400
MHz \cite{mgj+94}.  In these cases the assumption of a constant power
law is not correct and this results in an overestimated flux density at 325
MHz.

Navarro et al.\ \cite*{nbf+95} discovered that the flux density of PSR
J0218+4232 has a non-pulsed component. They find that the continuum
flux density at
325 MHz varies between 100 and 200 mJy. I used this flux density
estimate for a comparison with the WENSS flux density. An estimate from the
pulsed flux density was not derived.

\subsection{Source maps}

\begin{figure*}
\centerline{\psfig{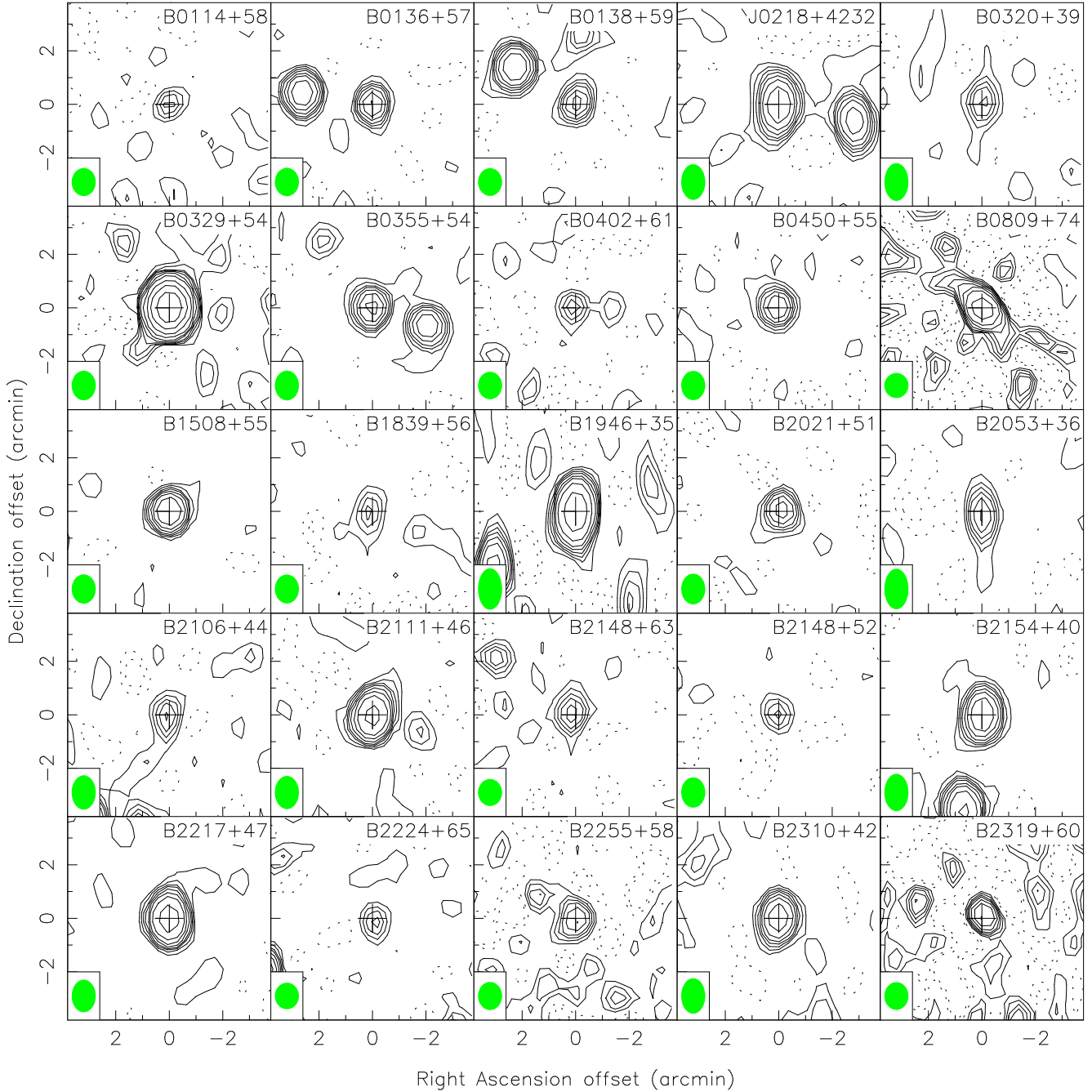}}
\caption{Maps of pulsars that are detected in the WENSS. The plots are centered
around the pulsar position, marked with a cross. The grey oval in the
lower left corner indicates the 
FWHM beam size. Contour levels are at 5, 10, 15, 20, 28, 40, 64, 128,
256, 512 and 1024 mJy (drawn lines) and at $-$5 and $-$10 mJy (dotted
lines).}
\label{fig:maps}
\end{figure*}

Fig.\ \ref{fig:maps} shows maps of the WENSS sources which are
correlated with a pulsar.  Most sources are consistent with a point
source convolved with the beam shape. The shapes of PSR~B0329+54 and
PSR~B0809+74 are affected by scintillation (see Sect.\
\ref{sec:discussion}). The integrated and peak flux densities of PSR~J0218+4232
differ by 20 percent. This suggests that the source is extended.
However, a two-dimensional fit shows that its
shape is not significantly different from the beam shape.

\section{Marginally detected pulsars}
\label{sec:marginal}

The WENSS source finding routines have a flux density threshold at five times
the local noise level ($5S_{\rm noise}$). Since in this paper single
trials are done to find a correlation at a very small number of
positions, a $3S_{\rm noise}$ source can be marked as a marginal
detection. However, source fitting is less acurate.  Therefore, a 5 by
5 pixel box (1.75\arcmin) centered around the pulsar position is
searched for the pixel with the maximum flux density. This pixel was marked as
a marginal detection, if its flux density was greater than $3S_{\rm noise}$.
The positional uncertainty for these marginal sources is estimated
using the same equation as for the fitted sources \cite{rtb+97}. The
positional error for these marginal sources is approximately
42\arcsec. Since a 1.75\arcmin~box is searched, no marginal detection
is missed.

\begin{figure*}
\centerline{\psfig{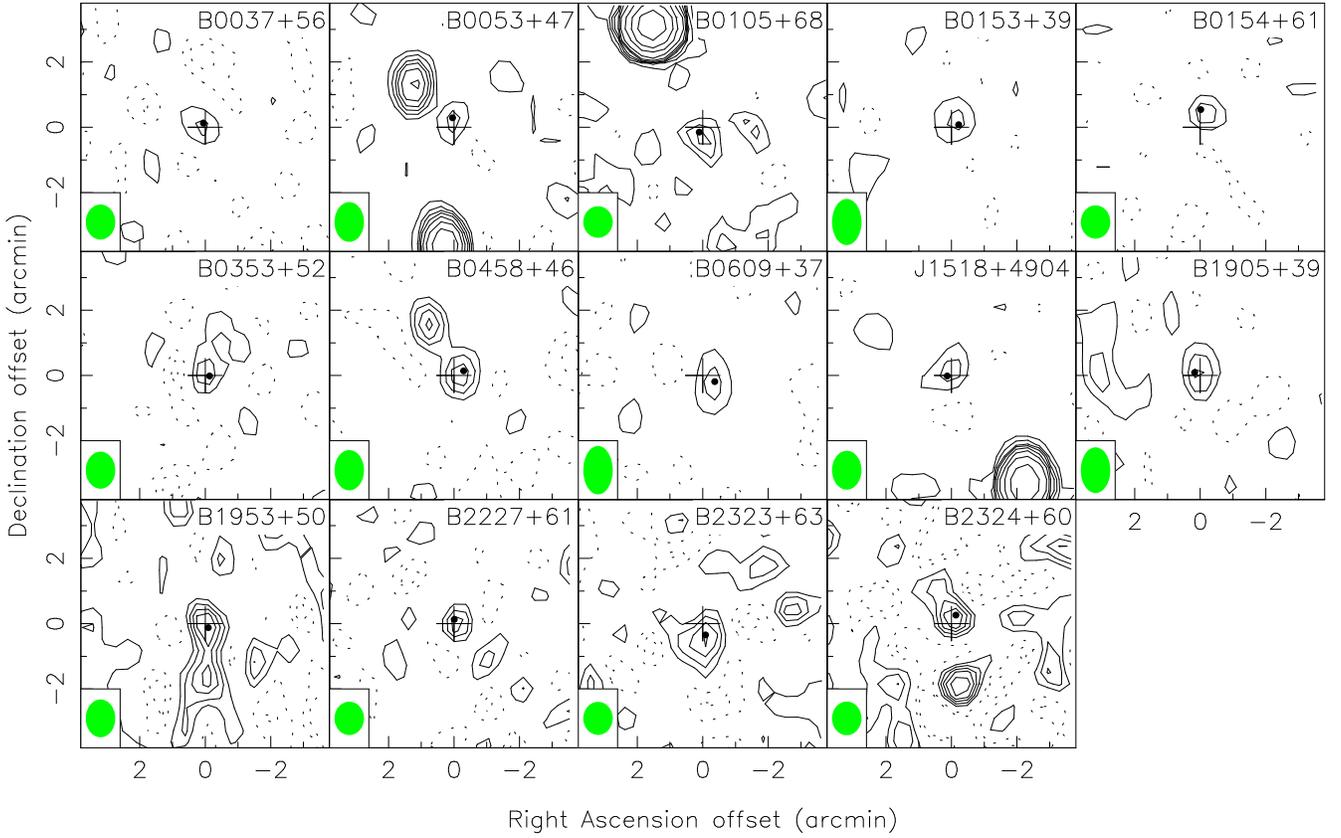}}
\caption{WENSS maps of the marginal detections which are correlated with a
pulsar. The plots are centered
around the pulsar position, marked with a cross. The dots indicate the maximum
pixel in a 5 $\times$ 5 pixel box around the centrum.
The grey oval in the lower left corner indicates the beam size. Contour
levels at 5, 10, 15, 20, 28, 40, 64, 128, 256, 512 and 1024 mJy (drawn
lines) and at $-$5 and $-$10 mJy (dotted lines).}
\label{fig:maps_marginal}
\end{figure*}

\begin{table}
\caption{List of estimated pulsar flux densities (PSR
$S_{325}$), WENSS maximum pixel flux densities (WENSS
$S_{325}$), the local noise level in the WENSS map ($S_{\rm noise}$)
and the ratio $R$ of the WENSS flux density and the noise level
for the 14 marginal detections.  PSR~B0053+47 and PSR~B0153+39 have
only upper limits for their spectral index, so they have a lower limit
for their 325 MHz flux density.  The estimate for 
PSR~J1518+4904 is based on its 370 MHz flux density (see text).
`D' in the last column indicates that a source near
the pulsar position was detected, although its expected flux
density was below three times the local noise level.}
\label{tab:flux_marginal}
\begin{tabular}{lr@{.}l@{ $\pm$ }r@{.}lr@{ $\pm$ }rrrl}
\hline\hline
\multicolumn{1}{c}{PSR} &
\multicolumn{4}{c}{PSR $S_{325}$} &
\multicolumn{2}{c}{WENSS} &
\multicolumn{1}{c}{$S_{\rm noise}$} &
R
\\
\multicolumn{1}{c}{name} &
\multicolumn{4}{c}{estimate} &
\multicolumn{2}{c}{$S_{325}$} & 
\\
\multicolumn{1}{c}{ } &
\multicolumn{4}{c}{(mJy)} &
\multicolumn{2}{c}{(mJy)} & 
\multicolumn{1}{c}{(mJy)} 
\\
%
%
\hline
~\\
B0037+56  &  10&3 &  0&7   & 11 &5 & 3.6 & 3.2 & D \\
B0053+47  &$>$ 4&2&  0&6   & 13 &5 & 3.5 & 3.7 & D~? \\
B0105+68  &   5&6 &  0&8   & 16 &5 & 3.5 & 4.6 & D \\   
B0153+39  &$>$ 6&7&  1&8   & 12 &5 & 3.5 & 3.4 & D~?\\   
B0154+61  &   8&4 &  1&1   & 14 &5 & 4.0 & 3.4 & D\\
~\\
B0353+52  &\m{19}&  &\m{3}&    & 14 &5 & 3.9 & 3.5 & \\
B0458+46  &  18&4 &  1&0   & 18 &6 & 4.6 & 4.0 & \\
B0609+37  &\m{21}&  &\m{3}&    & 13 &5 & 3.8 & 3.3 & \\ 
J1518+4904&\m{$\sim$ 8}&& \m{4}& & 13 &4 & 3.0 & 4.2 &  D~? \\
B1905+39  &\m{40}&  &\m{5}&    & 16 &5 & 3.8 & 4.2 & \\ 
~\\
B1953+50  &\m{37}&  &\m{6}&    & 29 &9 & 6.5 & 4.4 & \\  
B2227+61  &  28&6 &  1&7   & 18 &6 & 4.7 & 3.8 & \\   
B2323+63  &\m{11}&  &\m{3}&    & 21 &8 & 6.0 & 3.5 & D\\
B2324+60  &  21&7 &  1&8   & 27 &12& 8.9 & 3.0 & D\\ 
~\\
\hline\hline
\end{tabular}

\end{table}

The probability of finding a $3S_{\rm noise}$ (or higher) pixel in a
map with Gaussian noise is $2.7\cdot10^{-3}$. About 5 WENSS FWHM beams
fit into a 5 $\times$ 5 pixel box. Fifty-one boxes were
searched. Effectively, 255 trials have been performed. The binomial
probability that one $3S_{\rm noise}$ pixel is found is 0.35, the
probability that two are found is 0.12 and that three are found is
0.03. Therefore, there is a 50 percent probability that (at least) one
of our 14 marginal detection is just a noise fluctuation.

The maps of the marginal detections are shown in Fig.\
\ref{fig:maps_marginal}. The location of the pixel with maximum
flux density is
also indicated. The shapes of these sources is not as point-like as 
the strong detections with $S > 5 S_{\rm noise}$. The flux densities of the
marginal detections are listed in Table \ref{tab:flux_marginal},
together with an estimate of the pulsar flux density based on a similar
extrapolation of the pulsar spectrum as done in Sect.\
\ref{sec:flux}. The estimate for PSR~J1518+4904 is based on its
measured flux density at 370 MHz \cite{snt97}, since its spectrum as
plotted by Kramer et al.\ \cite*{kll+99} shows evidence for a
low frequency turnover.

The ratios of the extrapolated pulsar flux density and the WENSS
source flux density
are displayed in Fig.\ \ref{fig:flux-WENSS}. The spread is of the
order of a factor 1.5, which is comparable with the spread for the
detected sources that were discussed in the previous section.  Five
sources were detected, although their expected flux density was below three
times the local noise level (see Table \ref{tab:flux_marginal}). The
flux density at 325 MHz could not be estimated for three other pulsars,
since no reliable flux density data at other frequencies were available.

\begin{figure*}
\centerline{\psfig{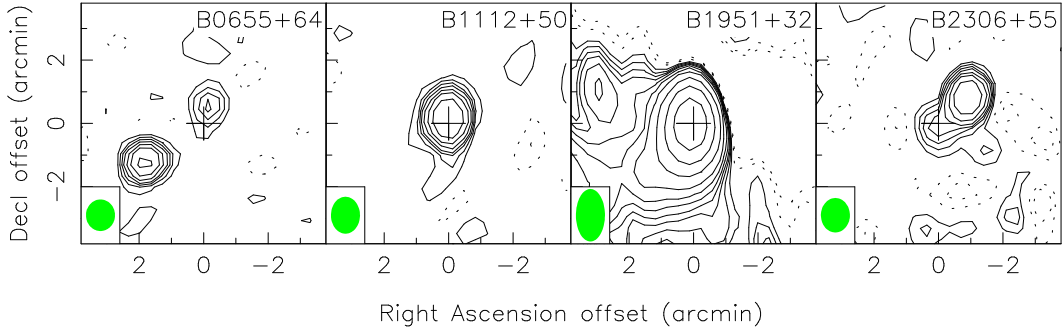}}
\caption{WENSS maps of the pulsars, which are confused by other radio
sources. The plots are centered
around the pulsar position, marked with a cross. The grey oval in the
lower left corner indicates the beam size. 
Contour levels at 5, 10, 15, 20, 28, 40, 64, 128, 256, 512 and 1024 mJy
(drawn lines) and at $-$5 and $-$10 mJy (dotted lines).}
\label{fig:maps_confusion}
\end{figure*}

The contours of four pulsars are confused by nearby radio
sources. These sources are discussed in the following and are
shown in Fig.\ \ref{fig:maps_confusion}.

\emph{PSR~B0655+64:} the WENSS source and the pulsar position are
4.3$\sigma$ apart. The estimated flux density of the pulsar at 325 MHz is 7
$\pm$ 2 mJy, but the WENSS source is 23 $\pm$ 5 mJy. Also, the NVSS
(see Sect.\ \ref{sec:discussion}) shows a radio source at the WENSS
position and clearly away from the pulsar position. Its flux density at 1400
MHz is about 5.6 mJy, while the pulsar flux density is expected to be 0.3
$\pm$ 0.1 mJy. The pulsar has a proper motion, but it is small and
directed towards negative declinations.

\emph{PSR~B1112+50:} The WENSS source is bright (135 mJy) and has an
accurate position. The separation between the fitted WENSS position
and the known pulsar position is 4.6$\sigma$. Extrapolation of the
pulsar spectrum results in an estimated flux density at 325 MHz of 16 $\pm$ 3
mJy, much less than that of the WENSS source.  Kaplan et al.\
\cite*{kcac98} and Han \& Tian \cite*{ht99} searched the NVSS for
pulsar counterparts and also noted that the pulsar is confused by a
strong NVSS source 12\arcsec~away.

\emph{PSR~B1951+32:} The coincident source in the WENSS catalog is
marked as extended and the emission is dominated by the supernova
remnant CTB80. The pulsar is associated with this remnant (Strom 1987,
Kulkarni et al.\ 1988). \nocite{str87} \nocite{kcb+88} The WENSS peak
flux density is 983 mJy, which is about a factor 70 stronger than the expected
pulsar flux density.

\emph{PSR~B2306+55:} It can be clearly seen in the map that this
source has two components, of which the weaker one is probably the
counterpart to the pulsar. This component is not listed in the WENSS
source list. The estimated pulsar flux density is about 30 $\pm$ 3 mJy. The
bright component of the WENSS source is 125 $\pm$ 9 mJy, separated
28\,$\sigma$ from the pulsar position. The second component has
a flux density of approximately 24 mJy.

In all these four cases I conclude that the source in the WENSS catalog
and the pulsar are unrelated.
Galama et al.\ \cite*{gbp+97} reached the same conclusion for PSR~B0655+64.

\section{Non-detected pulsars}
\label{sec:nondetections}

\begin{table}
\caption{List of the pulsars with no WENSS counterpart, their expected
flux densities at 325 MHz and the local noise in the WENSS map.  The expected
flux density for pulsars B0841+80, J1012+5307 and B1639+36A is equal to their
400 MHz flux density. The flux density of
PSR~J2044+46 is not known. `ND' in the
last column indicates, that the expected pulsar flux density is higher than
three times the local noise level.}
\label{tab:flux_nondetections}
\begin{tabular}{lr@{\mbox{}}l@{ $\pm$ }r@{\mbox{}}lrr}
\hline\hline
\multicolumn{1}{c}{PSR} & \multicolumn{4}{c}{PSR $S_{325}$} & 
                          \multicolumn{1}{c}{$S_{\rm noise}$} \\
\multicolumn{1}{c}{name}& \multicolumn{4}{c}{estimate} & \\
~                       & \multicolumn{4}{c}{(mJy)} &
                          \multicolumn{1}{c}{(mJy)} \\
\hline
~\\
B0011+47  & 19&   &  5&   & 3.3  & ND\\
B0045+33  &$>$ 3&.5& 0&.5 & 3.5  & ?\\
B0052+51  &  4&.1 &  0&.5 & 3.9  \\
B0059+65  & 11&.4 &  1&.3 & 3.2  & ND \\
B0105+65  & 14&.6 &  1&.9 & 3.2  & ND \\
~\\
B0144+59  &  7&.0 &  0&.5 & 3.4  \\
B0226+70  &  3&.5 &  0&.4 & 3.5  \\
B0331+45  & 14&   &  3&   & 4.2  & ND \\
B0339+53  &  4&.9 &  1&.0 & 4.0  \\
B0410+69  & 10&.5 &  0&.7 & 4.2  \\
~\\
B0643+80  & 11&.2 &  1&.8 & 1.9  & ND \\
B0751+32  & 12&.4 &  2&.5 & 3.5  & ND \\
B0841+80  &  1&\multicolumn{3}{@{\mbox{}}l}{.5} & 2.1  \\
B0917+63  &  6&.9 &  1&.4 & 3.5  \\
J1012+5307& 30&   & 30&  & 4.2 & ND \\
~\\
B1322+83  & 15&   &  7&   & 2.2  & ND \\ 
B1639+36A &  3&\multicolumn{3}{@{\mbox{}}l}{.0} & 4.3  \\
B1753+52  &  5&.1 &  1&.1 & 3.8  \\
B1811+40  & 18&.4 &  1&.0 & 3.3  & ND \\
B2000+32  &  8&.3 &  0&.5 & 7.3  \\
~\\
B2002+31  & 27&.9 &  1&.8 & 7.8  & ND \\
B2011+38  & 34&.3 &  1&.8 & 8.5  & ND \\
B2022+50  &  7&.1 &  0&.8 & 4.8  \\
B2027+37  & 27&   &  3&   & 7.4  & ND \\
B2035+36  &  8&.7 &  2&.2 & 7.0  \\
~\\
B2036+53  &  4&.6 &  0&.8 & 5.4  \\
J2044+46  & \multicolumn{4}{c}{ }  & 6.4 & ? \\
B2045+56  &  6&.6 &  1&.8 & 4.8  \\
B2241+69  &  3&.6 &  1&.4 & 4.2  \\
B2303+30  & 38&   &  4&   & 3.7  & ND \\
~\\
B2303+46  &  1&.7 &  0&.7 & 3.8  \\
B2334+61  & 16&   &  4&   & 6.9  \\
B2351+61  & 23&   &  4&   & 5.5  & ND \\
~\\
\hline\hline
\end{tabular}

\end{table}

Table \ref{tab:flux_nondetections} lists the pulsars that have no
counterpart in the WENSS.  In 14 cases the expected pulsar flux density is
higher than three times the local noise level. Still, the pulsar was
not detected. In two cases no reliable pulsar flux density estimate at 325 MHz
is available. In three other cases the estimate is based on 400 MHz
observations. In case of PSRs~B0841+80 and B1839+36A this was done,
because there was no spectral information available. The spectrum of
PSR~J1012+5307 might also have a low frequency turnover (see its
spectrum as plotted by Kramer et al.\ 1999) \nocite{kll+99}. Its flux density is
known to vary by up to a factor four from its mean value of 30 mJy \cite{nll+95}.

Five (and possibly eight) pulsars are detected, which
were expected to be not detectable. The number of non-detected
sources that were expected to have a flux density greater than the detection
limit, should be roughly the same as the number of unexpected detections.
The difference may be due to Poisson
fluctuations in the (small) number of pulsars in this study.

\section{Sources in the NVSS}
\label{sec:specind}

\begin{table}
\caption{List of spectral indices from Lorimer et al.\ (1995) and from
a WENSS-NVSS flux density comparison for all pulsars with both a WENSS and a
NVSS counterpart. The spectral index for PSR~J1518+4904 was taken from
Kramer et al.\ (1999). The errors on the pulsar spectral index (second
column) include variations due to scintillation, the errors on
the WENSS-NVSS spectral index do not include scintillation effects.}
\label{tab:specind}
\begin{tabular}{ll@{ $\pm$ }ll@{ $\pm$ }l}
\hline\hline
\multicolumn{1}{c}{PSR} &
\multicolumn{2}{c}{Spec.ind.} &
\multicolumn{2}{c}{Spec.ind.} \\
\multicolumn{1}{c}{name} &
\multicolumn{2}{c}{Lorimer}  &
\multicolumn{2}{l}{WENSS--} \\
\multicolumn{1}{c}{ } &
\multicolumn{2}{c}{et al.} &
\multicolumn{2}{r}{NVSS} \\
\hline
~\\
B0136+57  & $-$1.3 & 0.1 & $-$1.96 & 0.10 \\
B0138+59  & $-$1.8 & 0.2 & $-$1.54 & 0.09 \\
B0329+54  & $-$1.6 & 0.2 & $-$1.35 & 0.04\\
B0353+52  & $-$1.5 & 0.1 & $-$0.99 & 0.25\\
B0355+54  & $-$0.9 & 0.2 & $-$1.41 & 0.06\\
~\\
B0402+61  & $-$1.4 & 0.1 & $-$1.36 & 0.19\\
B0450+55  & $-$1.3 & 0.2 & $-$1.45 & 0.07\\
B0809+74  & $-$1.7 & 0.2 & $-$2.27 & 0.07\\
B1508+55  & $-$2.3 & 0.1 & $-$1.95 & 0.05\\
J1518+4904& $-$1.5 & 0.2 & $-$0.72 & 0.20 \\
~\\
B1839+56  & $-$1.4 & 0.2 & $-$0.57 & 0.15\\
B1946+35  & $-$2.2 & 0.1 & $-$2.57 & 0.07\\
B2021+51  & $-$0.7 & 0.3 & $-$0.79 & 0.08\\
B2106+44  & $-$1.3 & 0.1 & $-$1.31 & 0.13\\
B2111+46  & $-$2.0 & 0.1 & $-$1.05 & 0.04\\
~\\
B2154+40  & $-$1.5 & 0.1 & $-$1.16 & 0.04\\
B2217+47  & $-$2.8 & 0.3 & $-$2.75 & 0.10\\
B2224+65  & $-$1.8 & 0.1 & $-$1.12 & 0.16\\
B2310+42  & $-$1.5 & 0.2 & $-$1.44 & 0.06\\
~\\
\hline\hline
\end{tabular}

\end{table}
\nocite{lylg95}

The NRAO VLA Sky Survey (NVSS, Condon et al.\ 1998) \nocite{ccg+98} is
a survey of the sky above declination $-$40\degr~at 1400~MHz. Its final
resolution is 45\arcsec~and is comparible with the WENSS resolution of
54\arcsec~(in right ascension). The bandwidth was effectively 42 MHz. 
The noise in the final NVSS maps is
about 0.45 mJy. These maps are constructed by taking the average of
a number of snapshot observations. Each point on the sky is observed in
about three snapshots. 

Seventeen of the twenty-five pulsars with a WENSS counterpart are also detected in the
NVSS (see Kaplan et al.\ \cite*{kcac98}, who did not include PSR B0138+59 and Han \&
Tian \cite*{ht99}). Two pulsars (PSR~B0353+52 and PSR~J1518+4904) with a marginal WENSS
counterpart coincide with a NVSS source. Kaplan et al.\ \cite*{kcac98} also lists the
WENSS sources which coincide with their NVSS sources. Besides the three pulsars already
mentioned, PSR~B0809+74 is also not in their list, though it matches their coincidence
criterion. Eight pulsars with a WENSS counterpart are not detected in the NVSS. This
can be caused by scintillation or these pulsars must have a steep spectrum (spectral
index between about $-$1.6 and $-$2.0). Two pulsars (PSRs~J0218+4232 and B2319+60) may
have a very steep spectrum (spectral index less than $-$2.6 and $-$2.3, respectively).

From the WENSS and NVSS flux density values the spectral index can be calculated.
These can be compared with the spectral indices as
determined from a large number of dedicated pulsar observations, as by
LYLG. The latter indices will suffer less from scintillation effects. The
uncertainty due to scintillation is included in the errors given by LYLG.

\begin{figure}
\centerline{\psfig{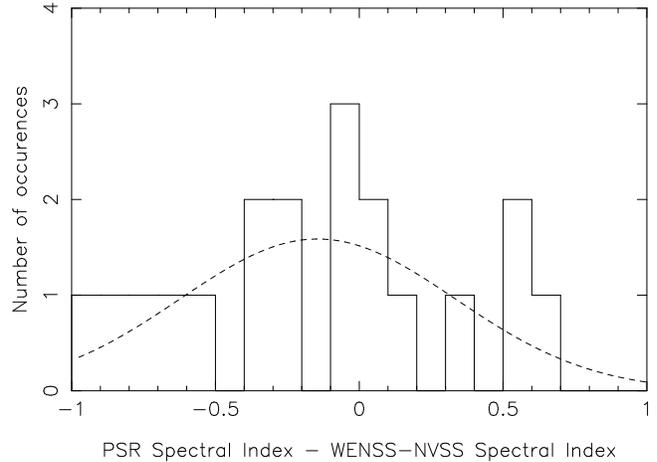}}
\caption{Histogram of the differences of the spectral index determined from the
WENSS and NVSS source flux densities and of the pulsar spectral index, as
determined by a large number of dedicated pulsar observations (Lorimer et
al.\ 1995). The dashed line is a Gaussian distribution with the same mean
and standard deviation as the histogram.}
\label{fig:specind}
\end{figure}

From Table \ref{tab:specind} it is clear that there are some large
differences between
the spectral index as determined by the WENSS and NVSS flux densities and the
long term averaged spectral index. The differences are displayed in Fig.\
\ref{fig:specind}. This histogram has a mean of $-$0.14 and a standard
deviation of 0.48. There is a large spread and a WENSS-NVSS source
flux density comparison is not a very precise way to determine a spectral index. 
This will limit the success of a
pulsar candidate selection based on their WENSS-NVSS spectral index.

\section{Discussion}
\label{sec:discussion}

The deviations between the flux density values of pulsars and the correlated
WENSS and NVSS sources are mainly caused by scintillation
effects. Small-scale inhomogeneities in the interstellar medium affect
the travel path of radio waves and can amplify or weaken them. For
reviews, see Rickett \cite*{ric90} or Narayan \cite*{nar92}. Walker
\cite*{wal98} summarizes the involved equations and
dependencies. 

In the case of {\sl strong} scintillation, the phase changes due to
the scattering in a certain region are larger than the changes in
phase due to normal geometry.  The scintillation is called {\sl weak}
in the opposite case. The boundary between the two is dependent on the
frequency of the radio waves and the distance to the pulsar.  At 325
and 1400 MHz almost all pulsars are in the strong scintillation
regime.

Two types of strong scintillation exist: diffractive and refractive.
They differ in their typical timescale $\tau$, frequency bandwidth
$\Delta \nu$ and size of the resulting flux density variation. This strength of
the scintillation is usually quantified by the modulation index, i.e.\
the rms fractional flux density variation.
 
In the case of strong scintillation waves from multiple locations in
the scattering region interfere constructively (or destructively). Both
the typical timescale and the frequency bandwidth are small. The
modulation index equals one. The timescale is dependent on the
relative velocities of the pulsar, the interstellar medium and the
Earth.  For the WENSS observations of pulsars typical timescales are 1
to 10 minutes and typical bandwidths are 10 Hz to 500 kHz.  This means
that for almost all pulsars any variations due to diffractive
scintillation are averaged out over the 5 MHz WENSS bandwidth and when
the 6 $\times$ 18 short observations spread over 6 $\times$ 12 hours
are combined.

\begin{figure}
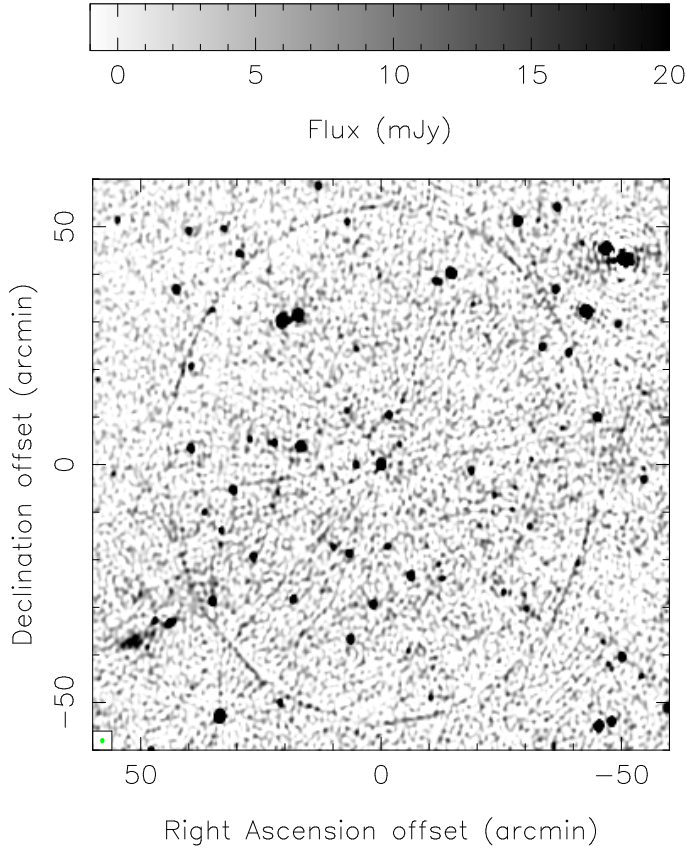

\centerline{\hspace*{0.13\columnwidth}
            \psfig{figure=0329_scale.ps,angle=0,width=0.9\columnwidth,clip=t}}
\vspace*{4mm}
\centerline{\psfig{figure=0329.ps,angle=0,width=\columnwidth,clip=t}}
\caption{Greyscale WENSS map of PSR~B0329+54. The plot is centered
around the pulsar position. To improve the visibility of the spokes,
the greyscale range has been adjusted (top panel).
The grey oval in the lower left corner indicates the
FWHM beam size. The spokes are caused by diffractive scintillation,
the ring is caused by refractive scintillation.}
\label{fig:0329}
\end{figure}

Only for PSR~B0809+74 are the diffractive scintillation timescale and
bandwidth large enough that some effect remains. From the equations
given by Walker \cite*{wal98} one finds a timescale of 12 minutes and a
bandwidth of 500 kHz. The actual value of the typical
scintillation timescale and bandwidth are even higher, since several
authors have already shown that the Taylor-Cordes distance model
\cite{tc93} gives too small predictions for this pulsar (e.g. Rickett
et al.\ 2000). \nocite{rcm00} Diffractive scintillation can explain
the WENSS image of this pulsar (Fig.\ \ref{fig:maps}). As 40 minutes
is the time between two observations of a field in one mosaic, flux density
variations on that time scale cause spokes in the map. Such spokes
can also be seen in the complete map of PSR~B0329+54 (Fig.\
\ref{fig:0329}).

Refractive scintillation is caused by the focussing effect  of a large
scattering region. The timescales and bandwidths involved are much
larger than in the case of diffractive scintillation. The modulation
index is also smaller. The scintillation bandwidth is of the order of the observing
frequency. Refractive time scales for the pulsars detected in
the WENSS vary from a couple of days to several years and the expected
modulation index from 0.05 to 0.3. 

If the refractive time scale is less than the time between
observations of the same mosaic (couple of days to several years), 
any flux density variation will be averaged
out when the mosaics are combined. However, stong flux density variations
between 12 hour sessions cause a ring at the first grating
ring of the synthesised beam. Since the mutual distances between the
dishes are multiples of 72 m, the ring will have radius of 72 m /
($c/$325~MHz) radians, i.e.\ 44\arcmin~in right ascension and
44\arcmin~ $\times ~\cosec ~\delta$ in declination. The second and higher grating
rings are not visible, since data far from the field center gets a low
weighting factor when the final image is created.

It is hard to give good estimates for the expected diffractive and
refractive scintillation timescales. They
depend on the often poorly known pulsar velocity.
For large pulsar velocities (compared to the velocity of the
interstellar medium, being about 50 km/s) the dependence is as one over
the square root of this velocity. Since some
pulsar velocities might be up to several hundred kilometers per
second, this cannot be neglected. I divided the pulsars that are
detected in the WENSS in two groups, based on their expected
refractive modulation index if their velocity is neglected.
Both groups had similar relative deviations between their measured and
expected flux densities.

The observed modulation index is about 0.4, 
much larger than the expected value of about 0.2. The WENSS pulsar
flux densities
vary more due to refractive scintillation than predicted by the equations.
This has been observed before
by several authors and is attributed to the assumption that
the turbulence in the interstellar medium has a Kolmogorov
spectrum (e.g.\ Blandford \& Narayan, 1985). \nocite{bn85}

The NVSS flux densities are even more affected by scintillation effects. Each
point in the NVSS maps is an average of about three snapshots.  Two of
these three are taken right after each other (snapshot series are
taken at constant declination and increasing right ascension, see
figure 7 in Condon et al.\ 1998) and very little averaging takes
place. The expected refractive modulation index at 1400 MHz is larger
than at the WENSS frequency. The expected diffractive frequency
bandwidth is also larger at the NVSS frequency, even relative to the
increased total bandwidth used in the NVSS. One therefore expects the
differences between the pulsar flux densities in the NVSS and
the flux densities
reported by LYLG to be larger than the differences reported in this
study. This is indeed observed: the modulation index of the WENSS
sources in Table \ref{tab:flux}, excluding PSRs J0218+4232, B0329+54
and B2021+51 is 0.40, the modulation index of the NVSS sources in
table 1 of Han \& Tian \cite*{ht99} is 0.60.

De Breuck et al.\ \cite*{bbrm00} show that the total spectral index
distribution of WENSS sources that are correlated with a NVSS source
($S_{\rm 325} > 50$ mJy and $\mid \! b \! \mid > 15^\circ$) has a mean of
$-$0.80 and a standard deviation of 0.24. Pulsar have a mean spectral index
of $-$1.6 (see LYLG).  An increase of 0.5 due to scintillation will move the
pulsar spectral index well into the distribution of normal sources, which
are much more frequent. The spectrum of some point-like quasars will
also be affected by scintillation and some of them will seem to have a
much steeper spectrum than they really have.  
This effect should be taken into consideration if pulsar
candidates are selected on the basis of their spectral index derived
from the WENSS and NVSS.

\begin{acknowledgements}
I thank F. Verbunt and A. G. de Bruyn for discussion and
comments. I thank B. W. Stappers for comments on the manuscript.
I am supported by The Netherlands Research School for 
Astronomy (NOVA), a national association of astronomy 
departments at the Universities of Amsterdam, Groningen, Leiden and Utrecht.
\end{acknowledgements}

\end{document}